\documentclass{article}

\usepackage[pdftex]{graphicx}
\usepackage{graphics}
\usepackage{subfigure}
\usepackage{color}
\usepackage{enumitem}
\usepackage{cite}
\usepackage{url}
\usepackage{makecell} 
\usepackage{booktabs} 

\newcommand\Mark[1]{\textsuperscript#1}

\par

\begin{document}
\title{Benchmarking Web API Quality -- Revisited}
\author{David Bermbach\Mark{1}, Erik Wittern\Mark{2}\\
\Mark{1}TU Berlin \& Einstein Center Digital Future\\Mobile Cloud Computing Research Group\\\texttt{db@mcc.tu-berlin.de}\\
\Mark{2}IBM, Hybrid Cloud Integration\\\texttt{erik.wittern@ibm.com}}

\maketitle

\subsection*{Abstract}

Modern applications increasingly interact with web APIs -- reusable components, deployed and operated outside the application, and accessed over the network. Their existence, arguably, spurs application innovations, making it easy to integrate data or functionalities.
While previous work has analyzed the ecosystem of web APIs and their design, little is known about web API quality at runtime. This gap is critical, as qualities including availability, latency, or provider security preferences can severely impact applications and user experience.

In this paper, we revisit a 3-month, geo-distributed benchmark of popular web APIs, originally performed in 2015. We repeat this benchmark in 2018 and compare results from these two benchmarks regarding availability and latency. We furthermore introduce new results from assessing provider security preferences, collected both in 2015 and 2018, and results from our attempts to reach out to API providers with the results from our 2015 experiments.
Our extensive experiments show that web API qualities vary 1.) based on the geo-distribution of clients, 2.) during our individual experiments, and 3.) between the two experiments. Our findings provide evidence to foster the discussion around web API quality, and can act as a basis for the creation of tools and approaches to mitigate quality issues.

\section{Introduction}\label{sec:introduction}

Today, mobile, web, or even desktop applications regularly rely on third-party data or functionalities, which they consume through web Application Programming Interfaces (web APIs).
Web APIs provide these applications with otherwise inaccessible resources such as access to global social networks (e.g., web APIs provided by Twitter, Facebook, or LinkedIn), advanced machine-learning capabilities (e.g., web APIs provided by IBM Watson or Google Cloud AI), or complex transaction processing (e.g., web APIs by Stripe or PayPal for payment processing or the Flight Booking API).
Today, application developers can build on
\begin{itemize} [leftmargin=1em]
\item ubiquitous technologies, e.g, the Hypertext Transfer Protocol (HTTP) or Asynchronous JavaScript and XML (AJAX),
\item architectural styles, e.g., the Representational State Transfer (REST)~\cite{Fielding:2000},
\item de-facto standards for describing web APIs, e.g., the OpenAPI Specification\footnote{\url{https://www.openapis.org/}}, 
\item research results, e.g., from service-oriented computing, cloud computing, or mash-ups, and finally
\item a myriad of web API client libraries in any established programming language,
\end{itemize}

\noindent so that integrating web APIs with an application no longer poses a technological challenge. Therefore, we now see thousands of public APIs as well as applications using them~\cite{wittern2014graph}.

In consequence, though, application developers now heavily rely on third-party entities beyond their control sphere for core functionality of their applications. This can have impacts on applications' user experience. For example, erroneous integration of core application capabilities, e.g., a payment service, via APIs can impede end user experience~\cite{Aue:2018}.
High request latency can lead to slow application response times, which have been found to disrupt users' flow of thought or eventually cause loss of attention~\cite{Nielsen:1994}. A long-term experiment performed by Google showed that increasing response times for search results artificially from 100ms to 400ms did measurably decrease the average amount of searches performed by users~\cite{Brutlag:2009}. User experience and, hence, application reputation is thus directly affected by actions and non-actions of the API providers. As another example, APIs may be discontinued or changed without notice, thus disabling applications. Previous work found that mobile applications silently fail and, in cases, even crash when confronted with mutated (e.g., adapted or faulty) web API responses~\cite{Espinha:2015}.
In sum, web APIs often present themselves as black-boxes with volatile \emph{qualities} -- i.e., availability, latency, security, or usage limitations -- to clients.

Based on this observation, we published a study on web API quality~\cite{paper_bermbach_api_benchmarking} in 2016, in which we described our findings from an experiment of benchmarking 15 endpoints\footnote{We denote an endpoint to be the combination of a resource, identified by a URL, and an HTTP \emph{method} as proposed in~\cite{Wittern:2015}.} of diverse web APIs for three months from geo-distributed clients. Core results were that availability and latency highly depend on the geo-origin of requests as well as the protocol used (HTTP vs. HTTPS). For this paper, we extend, repeat, and correct our previous study\footnote{Where applicable, this paper reuses parts of the text material and figures from our previous publication~\cite{paper_bermbach_api_benchmarking}.}. Namely, we repeated all experiments in 2018, reanalyze old and new availability and latency measurements, analyze yet unpublished measurements on provider security preferences, and present the results of trying to contact the providers of all benchmarked APIs with the goal of (i) finding explanations for the observations from our 2016 paper and (ii) to assess availability and quality of customer service. We thus make the following contributions:

\begin{enumerate}
	\item We extend our previous measurement method to also collect information on Transport Layer Security (TLS) cipher suite preferences of providers.
	\item We report findings on how TLS cipher suite preferences of providers evolved over time in both our 2015 and 2018 measurements.
	\item We report detailed findings from analyzing and comparing our 2015 and 2018 latency and availability measurements.
	\item We discuss the results of reaching out to the providers of all benchmarked APIs about our original findings from 2015.
\end{enumerate}

It should be noted that our work takes the perspective of an application developer, i.e., we have little insight into what happens behind the scenes but instead report things observable in practice: While API implementations may change or APIs may be discontinued, this is not a limitation of our results but rather a finding in itself.

This paper is structured as follows: we present background on web APIs, possible failures when consuming them, and web API qualities in Section~\ref{sec:background}. We present the implementation of our benchmarking client and the design of our experiments -- both with regard to benchmarked API endpoints and experiment setup -- in Section~\ref{sec:expdesign}. In the following sections, we compare findings from the 2015 and 2018 experiments, addressing availability in Section~\ref{sec:availability}, performance in Section~\ref{sec:latency}, and provider security preferences in Section~\ref{sec:security}. We present findings from reaching out to API providers with our 2015 results in Section~\ref{sec:reaching_out}. We discuss our approach, findings, and implications for application developers in Section~\ref{sec:discussion}. After presenting related work in Section~\ref{sec:relwork}, we conclude in Section~\ref{sec:conclusion}.

\section{Background\label{sec:background}}

In this section, we give an overview of selected qualities in web APIs and discuss how they can be measured. For this purpose, we start with a description of individual steps in performing web API requests (Section~\ref{sec:quality_http}) and potential root causes of failures (Section~\ref{sec:quality_errors}). Afterwards, we characterize the qualities which we have studied for this paper (Section~\ref{sec:quality_qualities}).

\subsection{Interaction with Web APIs}
\label{sec:quality_http}
Web APIs expose \emph{data}, e.g., a user profile or an image file, and \emph{functionalities}, e.g., a payment process or the management of a virtual machine through a resource abstraction. This abstraction enables users to manipulate these resources without requiring insight into the underlying implementation. 

\begin{figure}[t]
  \centering
  \includegraphics[width=0.9\linewidth]{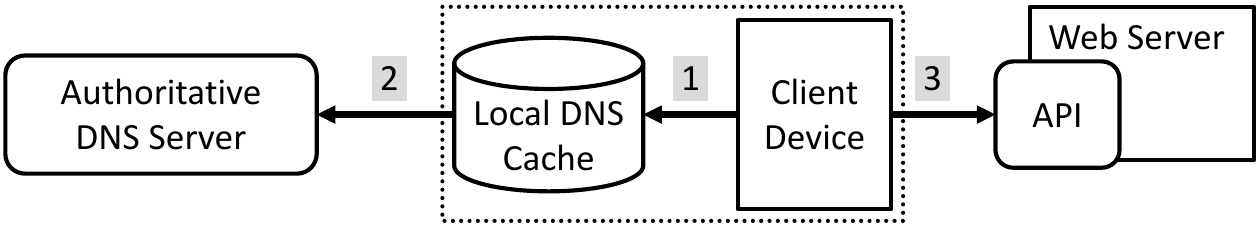}
  \caption{Overview of the Steps Involved in Sending an HTTP Request}
  \label{fig:http_request}
\end{figure}

Developers can access Web APIs through the \emph{Hypertext Transfer Protocol} (HTTP), which again uses the \emph{Transmission Control Protocol} (TCP) for error-free, complete, and ordered data transmission on the transport layer, and the \emph{Internet Protocol} (IP) at the network layer.
Figure~\ref{fig:http_request} illustrates the steps involved in a typical HTTP request\footnote{For simplicity's sake, we do not include possible complications such as proxies, keep alive connections, caches, or gateways in this Figure.}.

The resources exposed by an API are identified by \emph{unified resource locators} (URLs), describing the scheme to be used for interaction, the server Internet address, and the specific resource identifier.
The semantics of interactions with a resource depend upon the HTTP \emph{method}, e.g., GET, POST, or DELETE.
Before a client can send a request to the server that offers the web API, client and server need to establish a TCP connection.
For this purpose, the client first sends a lookup request for the URL of the server to a \emph{Domain Name Service} (DNS) server  which returns the IP address and port number of the target host.
If available, IP address and port may be returned from a local cache (step 1); otherwise, an external DNS authority is consulted (step 2).
Afterwards, the client opens a socket connection to the server, i.e., it initiates TCP's three-way handshake, thus, establishing a TCP connection (step 3).
Based on this connection, multiple HTTP requests with application data can be sent to the server.

If additional security is required, the client will typically use HTTPS which introduces the \emph{Transport Layer Security} (TLS) protocol\footnote{TLS has largely replaced its predecessor SSL which is typically supported only for compatibility with old clients.} between HTTP and TCP/IP. TLS has two main phases: a negotiation phase and a bulk data transfer phase. In the negotiation phase, the server authenticates itself through its X.509 certificate. Afterwards, the client sends its list of supported \emph{cipher suites} (a combination of symmetric encryption algorithm and a message authentication code (MAC)) to the server which then selects a cipher suite supported by both client and server and responds accordingly. Using asymmetric encryption (e.g., RSA) and key exchange protocols (e.g., DHE), client and server also agree on a symmetric key and other TLS sessions parameters.

After this TLS handshake has been completed, the server signals a change to the bulk data transfer phase. During that phase, each HTTP request is broken down into data packets which are -- based on the agreed session parameters -- encrypted and signed before transmission over the network. Cipher suite and protocol version determine whether encrypt-then-MAC or the reverse order is used. The recipient can then reassemble the original request and verify its integrity based on the received MAC.

\subsection{Sources of Failures}
\label{sec:quality_errors}
Considering the typical HTTP request flow described in Section~\ref{sec:quality_http}, a number of possible breakpoints emerge at which a request may fail~\cite{cito2015identifying}.
As we will see, while some of these are in control of a web API provider, others are not.

A \textbf{failed DNS lookup} is caused by attempting to look up a host for which no DNS entry exists or by a network partitioning which causes the lookup request to an authoritative DNS server to time out.
The first error source is rather unlikely for web API requests with the correct URL, as it would either imply the disappearance of the API's host altogether or indicate problems in the DNS system itself.
Typically, a failed lookup results in a timeout error reported to the client.
The second error source appears only in case that the network is not available and the DNS entry is not yet cached locally.

A \textbf{client connection error} appears if no TCP connection can be established between the client and the server hosting the web API.
Reasons for this error are network partitioning or that the server is in a state where it cannot accept connections (for example, because it crashed).

In the case of HTTPS, a request can also fail if authentication of the server is not possible due to certificate issues or if there is no cipher suite supported by both client and server.

A \textbf{client error} appears if the request sent by the client cannot be processed by the server.
One common reason for client errors is that the requested resource cannot be found on the server (resulting in a $404$ error code being returned).
Furthermore, users may not be authorized to access the requested resource.
The client may not have been aware of authentication mechanisms, e.g., basic authentication or OAuth, or may not own proper credentials.
Furthermore, providers may deny authorization for specific clients if their usage of an API exceeds certain thresholds.
A broad range of client errors are considered by HTTP and should result in the server sending 4xx status codes.
While these errors are attested to the client, it is important to note that their appearance can be tightly related to actions of the web API provider.
For example, many changes on the server, e.g., introducing authentication, removing or renaming resources, or changing request formatting, cause existing clients to malfunction, i.e., the client error is in fact caused by the web API provider.

A \textbf{server error} appears if the server fails to process an otherwise correct request.
Reasons for server errors may include failed lookups for resources in databases or errors in the execution of functionalities.
Server errors are, similar to client errors, considered by HTTP and should result in the server sending 5xx status codes.

\subsection{Qualities}
\label{sec:quality_qualities}
Systems have a number of properties. These can be functional, i.e., describe the abilities of said system, or non-functional, i.e., describe the quality of said system. Quality describes how ``good'' or ``bad'' a system fulfills its tasks along several dimensions\footnote{It depends on the respective quality what ``good'' or ``bad'' implies.} -- the qualities~\cite{book_bermbach_cloud_service_benchmarking}.

There is a plethora of qualities that we can see in web APIs. Examples range from availability and performance, to security, reliability, scalability, cost, or correctness (of results). All these qualities are inherently connected through complex direct and indirect tradeoff relationships~\cite{diss_bermbach}. In this paper, we focus on three qualities: availability, performance, and provider preferences regarding (transport layer) security.

\subsubsection{Availability}
Generally, availability describes the likelihood of a system -- here, a web API -- being able to respond to requests. Providing a concise definition of availability, though, is non-trivial: Does an API have to send correct responses or does it suffice if it is still able to tell about current problems? For this paper, we distinguish -- based on our previous work~\cite{paper_bermbach_api_benchmarking} -- three different kinds of availability to consider these questions:

\emph{Pingability} describes whether there is anything ``alive'' at the web API provider's site. This may be a load balancer or even a fault endpoint. For a single machine deployment, pingability describes whether said machine is reachable at an operating system level. Pingability is fulfilled if, at the web API's URL, some entity responds to basic low level requests, e.g., ping requests (using the ICMP protocol).

\emph{Accessibility} describes whether the resource represented by the web API is still accessible but not necessarily able to fulfill its task. For a single machine deployment, accessibility describes whether the web server component is reachable but does not require the hosted application logic to be accessible. A web API is accessible if it responds to HTTP requests using one of the predefined HTTP status codes.

\emph{Successability} describes whether the web API is fully functional. For a single host deployment, it requires the application logic to be working\footnote{Please note that successability does not say anything about correctness of results.}. Hence, we define successability to be fulfilled if a web API responds to requests using 2xx or 3xx status codes.

\subsubsection{Performance}
Performance has two dimensions: \emph{latency} and \emph{throughput}. Latency describes the amount of time between the start of a request at the client and the end of receiving a response, also at the client. Throughput, on the other hand, describes the number of requests a web API is handling at a given point in time. Typically, throughput measurements try to determine the maximum throughput, i.e., the maximum number of requests that a web API is able to handle without timeouts~\cite{book_bermbach_cloud_service_benchmarking,paper_kossmann_cloud_datastore_benchmarking,paper_kuhlenkamp_vldb,paper_klems_quality_measurement_framework}. 

Usually, these two dimensions are interconnected: If the load increases towards maximum throughput, then latency will increase. If this is not the case, then the system behind the web API is typically referred to as elastically scalable~\cite{paper_kossmann_cloud_datastore_benchmarking,paper_kuhlenkamp_vldb}.

\subsubsection{Security}
\label{sec:background_security}
Security is typically characterized along several dimensions -- here, we will focus on the two arguably most relevant for TLS: \emph{confidentiality} and \emph{integrity}. Basically, confidentiality describes whether unauthorized entities are able to access the content of API requests, and integrity asserts that transmitted data packets cannot be manipulated without the manipulation being noticed, cf., e.g.,~\cite{paper_mueller_tls_benchmarking,paper_pallas_security_performance_hbase}.

It is hard to quantify how secure a web API interaction is with regards to confidentiality and integrity as this would require knowledge on all possible attack vectors. However, it is possible to interpret the selected \emph{cipher suites} of actual API requests and the general preference order of the API provider and distinguish between weaker and stronger cipher suites. This is particularly important since the server, i.e., in this case the API provider, selects the cipher suite from the client's list of supported cipher suites.

Cipher suites list cryptographic algorithms for use in TLS (and previously SSL). A cipher suite generally names the algorithms to use for 1) key exchange, 2) authentication, 3) bulk encryption (this algorithm is also called the ``cipher''), and 4) message authentication code (MAC; which is used to ensure the integrity of a message). Servers typically support a ranked list of cipher suites, from which one suite to use is agreed-on with the client during the TLS handshake procedure. Using the preference list of provider cipher suites, we can rate the security preferences of the provider.

\section{Experiment Design\label{sec:expdesign}}

In this section, we describe our experiment design. We start by laying out the goals of our experiments (Section~\ref{sec:expdesign_goals}) before describing the qualities we want to measure (Section~\ref{sec:expdesign_qualities}). We state the set of API endpoints in focus, and describe required changes that we needed to make between our two experiment runs (Section~\ref{sec:expdesign_endpoints}). We then outline our measurement approach (Section~\ref{sec:expdesign_measurements}) before finally describing the implementation and deployment of our measurement client (Section~\ref{sec:expdesign_implementation}).

\subsection{Goals}
\label{sec:expdesign_goals}
In our experiments, we measure qualities of web APIs \emph{as perceived by applications}. Our goal is, for one, to be descriptive, i.e., we want to systematically assess the qualities that applications can expect from APIs. Furthermore, the goal of our measurements from an application perspective is to provide a basis for mitigation mechanisms that clients (i.e., applications) can use, as we summarize in Section~\ref{sec:discussion} based on our original paper~\cite{paper_bermbach_api_benchmarking}.

As such, we treat the API itself as a blackbox in our experiments, and measure qualities on the client level. This means that measured qualities may not only be affected by things in control of the API provider, e.g., the design or provisioning of API servers, but also by intermediaries, e.g., the network connection between client and API. As such, the results of our measurements cannot be used to compliment or blame individual API providers. Rather, they reveal real risks associated with varying or insufficient qualities that applications face when using APIs.

We initially measured web API qualities in 2015, and published a dedicated paper discussing our findings~\cite{paper_bermbach_api_benchmarking}. For this work, we repeated the same experiments once more in 2018 with the intention of comparing results and possibly revealing changes in qualities over time.

\subsection{Qualities in Scope}
\label{sec:expdesign_qualities}
Our experiments cover all three API qualities described in Section~\ref{sec:quality_qualities}, namely, availability, performance, and provider security preferences. We selected these qualities as they all have possibly significant implications for applications. The availability and performance of an application can be directly impacted by these qualities in used APIs. Both, availability~\cite{HenryMartinez:2009,Zambon:2011} and performance~\cite{Nielsen:1994,Brutlag:2009} have been shown to affect user satisfaction, user retention, and ultimately revenues. Implications of poor security choices, while harder to quantify, obviously impact the success of applications.

In addition to these technical qualities, we are interested in seeing how providers react to the results of our measurements. By trying to reach out to providers, we hope to get explanations for our results and to provide feedback from our side. Furthermore, we aim to explore how easy it is to reach providers, i.e., how easy it is for application developers to get help.

We deliberately refrain from measuring throughput (and, thus, also scalability since we do not have any insight into the API provider's implementation) as we do not intend to unreasonably strain APIs, and do not want to violate terms of service explicitly prohibiting excessive request rates.

\subsection{Selected Web API Endpoints}
\label{sec:expdesign_endpoints}
For our 2015 experiments, we originally selected $15$ unauthenticated API endpoints ($11$ accessible via HTTP or HTTPS, $3$ accessible only via HTTP, and one accessible via HTTPS only)~\cite{paper_bermbach_api_benchmarking}.\footnote{We rely on unauthenticated endpoints only as authentication mechanisms may require mandatory manual setup steps, may affect measurments by introducing additional roundtrips, or may necessitate the use of software development kits (SDKs), again, with possible runtime implications.} Aiming to compare experiment results, we relied on the same endpoints in our 2018 experiments. The selected endpoints stem from a broad variety of different providers with regards to company size, country of origin, local or global target users, public or private sector. We specifically included some of the most well-known providers, e.g., Google, Apple, Amazon, and Twitter. Table~\ref{table:apis} gives an overview of these web API endpoints, and of the protocols they supported.

\begin{table}[!t]
  \center
  \caption{Benchmarked API Endpoints and Supported Protocols in 2015~\cite{paper_bermbach_api_benchmarking}.}
  \begin{tabular}{ p{2.4cm}  c  p{6.3cm}  }
  \textbf{API Name}             & \textbf{\tiny{ICMP/HTTP/HTTPS}} & \textbf{Request Meaning} \\
  \toprule
  Apple iTunes         & X   / X   / X     & Get links to resources on artists \\
  Amazon S3            & -   / X   / X     & Get file list for the 1000 genomes public data set \\
  BBC                  & -   / X   / -     & Get the playlist for BBC Radio 1 \\
  Consumer Finance     & X   / X   / X     & Get consumer complaints on financial products\\
  Flickr               & X   / X   / X     & Get list of recent photo and video uploads \\
  Google Books         & X   / -   / X     & Get book metadata by ISBN \\
  Google Maps          & X   / X   / X     & Query location information by address \\
  MusicBrainz          & X   / X   / X     & Get information about artists and their music \\
  OpenWeather Map      & X   / X   / -     & Get weather data by address \\
  Postcodes.io         & X   / X   / X     & Get location information based on UK zip codes \\
  Police.uk.co         & -   / X   / X     & Get street level crime data from the UK \\
  Spotify              & X   / X   / X     & Get information on a given artist \\
  Twitter              & X   / X   / -     & Get the number of mentions for a given URL \\
  Wikipedia            & X   / X   / X     & Get a Wikipedia article \\
  Yahoo                & X   / X   / X     & Get weather data by address \\
  \bottomrule
  \end{tabular}
  \label{table:apis}
\end{table}

Already during our 2015 experiments, two endpoints became permanently unavailable~\cite{paper_bermbach_api_benchmarking}.

Since then there have been additional changes as some resources, their endpoints, or even entire APIs that we originally targeted were no longer accessible in 2018. Table~\ref{table:changes} lists all changes to the HTTP and HTTPS endpoints for the anonymized APIs. Notably, some HTTP endpoints available in 2015 have been deprecated since. In one case, for API 4, the originally targeted resource seized to exist, and we had to replace it (with an, unfortunately, much smaller one). Table~\ref{table:changes} also indicates changes in response size between 2015 and 2018. Response sizes evolved significantly, even for cases where the API, endpoint, or even targeted resource remained constant. For some APIs, we cannot report response size changes as we were no longer able to complete requests, whether with HTTP or HTTPS.

\begin{table}[!t]
  \center
	\vspace{0.3cm}
  \caption{Changes in API Endpoints between 2015 and 2018}
  \vspace{0.3cm}
	\begin{tabular}{ c  l  l  l }
\textbf{API} & \textbf{HTTP change} & \textbf{HTTPS change}  & \textbf{Resp. size} \\
\toprule                                           
1   & -                   & -                   & -3.1\%    \\ 
2   & no longer available & -                   & n.a.      \\ 
3   & -                   & -                   & +12.48\%  \\ 
4   & new resource        & new resource        & -99.92\%  \\ 
5   & -                   & -                   & +254.56\% \\ 
6   & -                   & -                   & -13.52\%  \\ 
7   & -                   & -                   & +25.57\%  \\ 
8   & -                   & -                   & -45.68\%  \\ 
9   & -                   & -                   & +42.85\%  \\ 
10  & -                   & -                   & +16.03\%  \\ 
11  & -                   & -                   & +50.03\%  \\ 
12  & no longer available & -                   & n.a.      \\ 
13  & no longer available & no longer available & n.a.      \\ 
14  & no longer available & -                   & n.a.      \\ 
15  & no longer available & -                   & +38.53\%  \\ 
\bottomrule
\end{tabular}
\label{table:changes}
\end{table}

As our measurements include factors beyond the control of API providers, and as we do not intend to discredit individual API providers, we anonymize results. For the remainder of this paper we refer to the API endpoints in focus as API-1 to API-15. There is no correlation between these identifiers and the order of API endpoints in Table~\ref{table:apis}. However, we use the same mapping as in our 2016 paper~\cite{paper_bermbach_api_benchmarking} for comparability. We will reveal the mapping information upon request if we are convinced that the information will not be used to discredit individual providers.

\subsection{Measurement Approach}
\label{sec:expdesign_measurements}
The qualities availability, performance, and provider security preferences are not constant, but evolve over time. As such, we opt for long-lasting experiments, where we repeatedly measure these qualities. This approach allows us a) to increase the confidence in results as (temporary) variance can be ruled out, and b) to obtain insights into the evolution of qualities. Specifically, both in 2015 and in 2018, we executed experiments for three months each (from August 20th to November 20th 2015, and from January 30th to April 30th 2018).

We measure availability of an API in terms of pingability, accessability, and successability as described in Section~\ref{sec:quality_qualities}. I.e., during the duration of our experiments, we repeatedly pinged the API endpoints in question and sent both HTTP and HTTPS requests to them (given their support of these protocols). We store the results of ping attempts, whether HTTP(S) requests resulted in a response from the server, and, if so, what HTTP status code was returned. To measure latency, we also measure the time between sending an HTTP(S) request from the client and receiving a response. We perform latency and availability measurements periodically every five minutes.

We measure the provider's security preferences by requesting the ordered list of supported cipher suites. We request an updated list of preferred cipher suites every twelve hours.

The above qualities of a web API can depend on the geographic region that a client is in. For example, latency depends on the geographic distance between client and API server. Or, availability and cipher suite preferences may differ as dedicated servers may respond to requests in different regions. As such, we measure qualities in a geo-distributed way, allowing us to compare results from different parts of the world.

Finally, we try to reach out to providers about our benchmarking results. We try to find contact details for all providers using Google search but also via contacts of contacts on LinkedIn. To each contact we found, we tried to send the same text including our original paper and deviated only where necessary due to the respective communication channel (e.g., on Twitter). We describe the results of this reaching out process -- including both observations on contact methods but also on outcomes of attempted contact -- in Section~\ref{sec:reaching_out}.

\subsection{Benchmark Implementation}
\label{sec:expdesign_implementation}
We implemented a benchmarking client consisting of three parts:
First, a Java-based custom benchmarking client, that uses the standard library's HttpURLConnection class to execute HTTP and HTTPS requests to the API endpoints. We scheduled the function to run every five minutes using standard Java thread scheduling. Results are appended to a local file.
Second, a bash script that uses the standard Linux ping implementation to repeatedly ping the API endpoints in scope. We invoked the script every five minutes using the Java ProcessBuilder class, parsed the results, and also appended them to a local file.
Third, a bash script that uses the cipherscan\footnote{\url{https://github.com/jvehent/cipherscan}} open-source tool to request the API servers' preferred cipher suites. We scheduled this script to run every twelve hours by invoking it via the ProcessBuilder class and appended results to a local file.

The benchmarking client is parametrized with a list of API endpoints to target. It then starts measuring qualities using the three implementation parts. The client selects different starting points per protocol and API to avoid interference. We further implemented an HTML dashboard that allows to check that the client runs without issues and to see intermediary results. We make the benchmarking client publicly available as open-source.\footnote{\url{https://github.com/dbermbach/web-api-bench}}

As in 2015, we deployed the benchmarking client on Elastic Compute Cloud (EC2) instances in the Amazon Web Services (AWS) cloud in the following seven regions: US East (Virginia), US West (Oregon), EU (Ireland), Asia (Singapore), Asia (Sydney), Asia (Tokyo), and South America (Sao Paulo).

After completing the benchmark runs, we copied created log files onto local workstations for data preprocessing and analysis.\footnote{The data collected in our experiments is available at\\ \url{https://github.com/ErikWittern/web-api-benchmarking-data}}

\section{Availability Findings\label{sec:availability}}

In this section, we give an overview of our findings regarding availability of web APIs. Reported results are based on our experiments from both 2015 and 2018.

\begin{figure}[t]
  \centering
  \includegraphics[width=0.9\columnwidth]{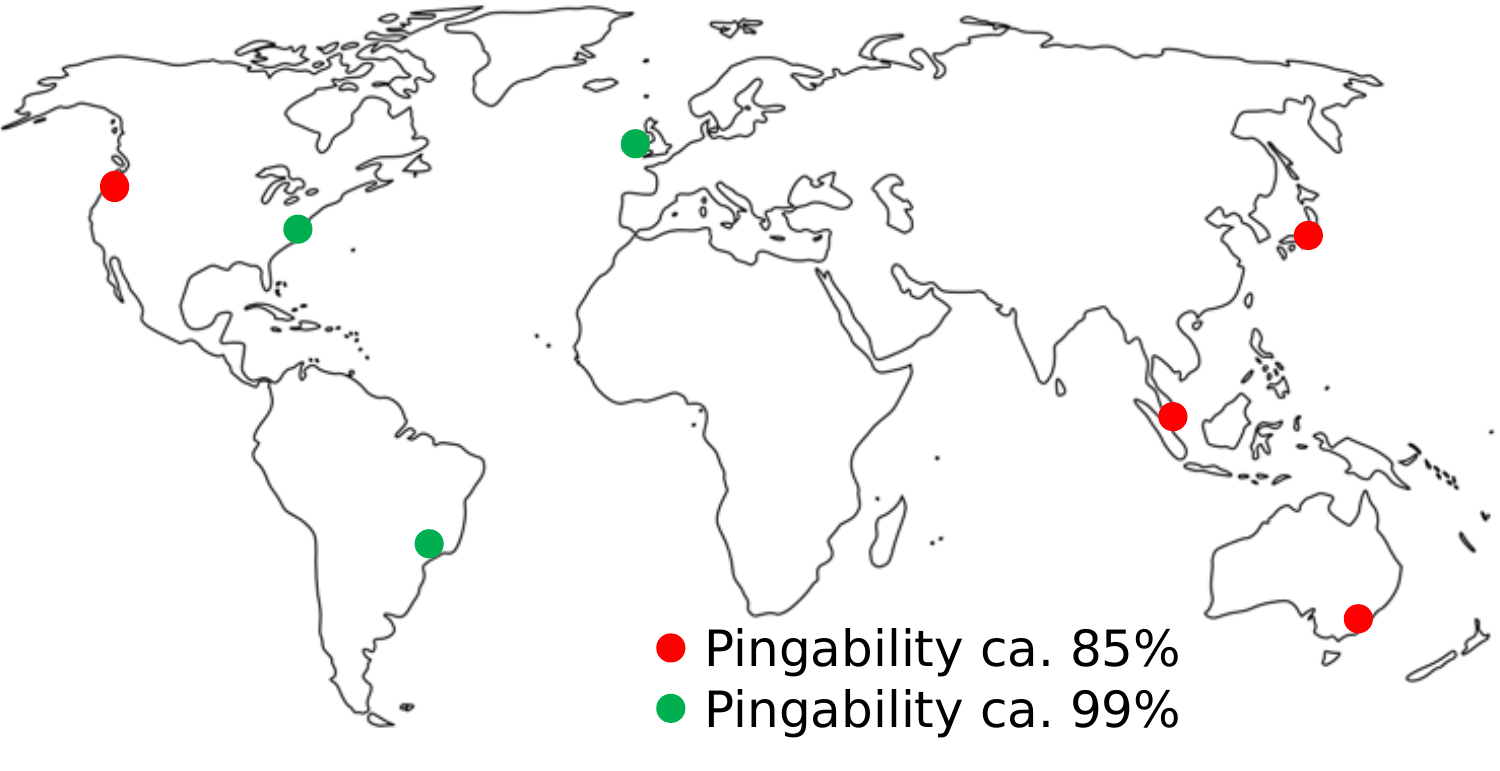}
  \caption{Pingability of API-6 in February 2018 by Region}
  \label{fig:ping-api6}
\end{figure}

\vspace{0.1cm}
\noindent
\textbf{Finding \#1: Pingability.} Overall, all APIs show a very high pingability of around 99\% in both 2015 and 2018. However, all APIs have very different pingability when comparing regions. This effect is particularly pronounced in the 2018 measurements for API-6: During the first month of our measurements, daily pingability went down to about 85\% for three out of seven regions while the remaining regions still had 99\% pingability. Figure~\ref{fig:ping-api6} shows the geographical distribution of these regions. Beyond API-6, this was observable for all other APIs as well though less extreme as for these APIs only the number of nines varied. Table~\ref{table:lost-pings} shows the distribution of lost ping packages across regions; overall, clients in South America appear to have a higher probability for pingability problems than elsewhere. Noticeably, API-6 which had the strongest variance and problems encountered their worst problem in Oregon and had very good pingability in Sao Paulo. All this, however, should be taken with a grain of salt as our experiment sent close to one million ping packages per API in total.

\begin{table}[t]
\center
\caption{Distribution of Lost Ping Packages Across Regions}
\begin{tabular}{ l  c  c   c  l  }
           
\textbf{API}     & \textbf{Best} & \textbf{Worst} & \textbf{Average} & \textbf{Worst Region} \\
\toprule
API-1             & 6 & 211 & 66 & Sao Paulo\\
API-3             & 4 & 374 & 82 & Sao Paulo\\
API-4             & 7 & 74 & 32 & Oregon\\
API-6             & 17 & 5852 & 2893 & Oregon\\
API-7             & 6 & 235 & 71 & Virginia\\
API-8             & 1 & 42 & 13 & Sao Paulo\\
API-9             & 20 & 476 & 155 & Sao Paulo\\
API-10             & 0 & 175 & 73 & Sao Paulo\\
API-15             & 0 & 31 & 13 & Sao Paulo\\
\bottomrule
\end{tabular}
\label{table:lost-pings}
\end{table}

\begin{figure}[t]
  \centering
  \includegraphics[width=\columnwidth]{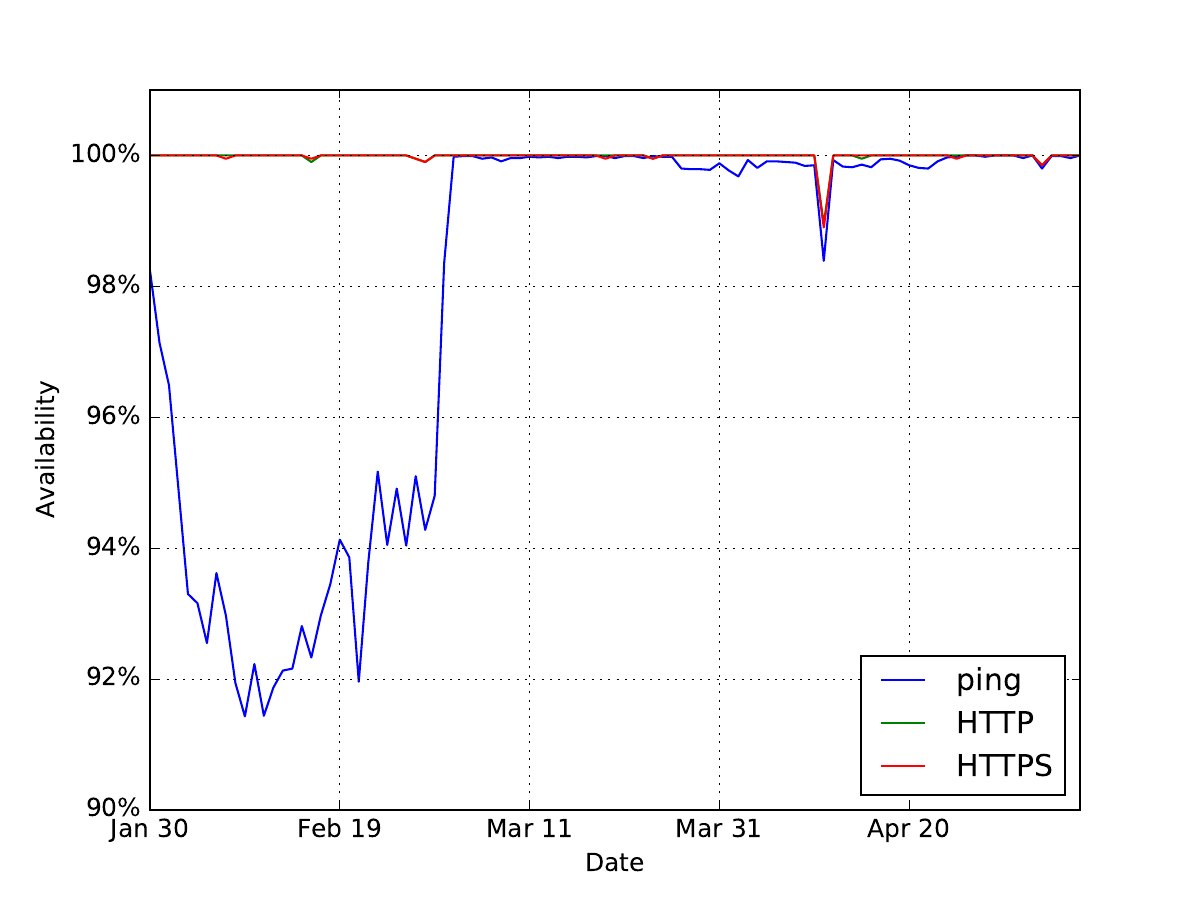}
  \caption{Comparison of Daily Pingability and Accessibility of API-6 in 2018}
  \label{fig:ping-vs-http-api6}
\end{figure}

\vspace{0.1cm}
\noindent
\textbf{Finding \#2: Pingability vs. Availability.} In our experiments, we also found that pingability is usually a good proxy for API availability. This, however, is not \emph{always} the case. Figure~\ref{fig:ping-vs-http-api6} shows the pingability and accessibility of API-6 in our 2018 benchmark. As can be seen in the left part, API-6 had severe pingability problems which were not related to the overall availability of the HTTP(S) endpoints. In this period of time, using pingability as a proxy for accessibility would severely underestimate the API accessibility. In contrast, the right half of the chart shows the expected correlation of pingability and accessibility.

\vspace{0.1cm}
\noindent
\textbf{Finding \#3: HTTP Status Codes.} Based on semantics of the HTTP protocol, applications are led to believe that they can expect 4xx HTTP status codes when they have done something wrong and 5xx when the fault lies with the server, or a response (with status codes 2xx or 3xx) otherwise. This could not be further from the truth. Across all our experiments, we found that up to nearly 90\% of all unavailabilities resulted in no response at all, i.e., a timeout or a loss of connection. See table~\ref{table:status-codes} which shows how frequently client-observable outcomes were a 4xx/5xx status code or some other sort of failure. Note, that during our 2015 benchmark, we saw two endpoints permanently going offline (the providers still exist at the time of writing this paper but deprecated the respective API endpoints): One permanently switched from 2xx to 4xx codes, the other did this for 48 hours only. Afterwards, requests failed without returning any status codes. In table~\ref{table:status-codes}, we show the full 2015 datasets as well as the dataset without these two APIs as both together account for about 94\% of all failed requests in 2015. Overall, we can only recommend that applications not solely rely on receiving HTTP status codes, but also explicitly handle timeouts or similar failures.

\begin{table}[!t]
\center
\caption{Distribution of Client-Observable Results (HTTP Status Code or None) for Failed Requests}
\begin{tabular}{ p{2.3cm}  l  r  r   r  }           
\textbf{Dataset} & \textbf{Protocol} & \textbf{4xx} & \textbf{5xx} & \textbf{None} \\
\toprule
2015 (all) & HTTP  & 43\% & 5\% & 53\%\\
 & HTTPS & 3\% & 8\% & 89\% \\
\midrule
2015 (w/o & HTTP & 0\% & 74\% & 26\% \\
offline endpoints) & HTTPS & 0\% & 89\% & 11\% \\
\midrule
2018 & HTTP & 0\% & 87\% & 13\% \\
 & HTTPS & 0\% & 80\% & 20\% \\
\bottomrule
\end{tabular}
\label{table:status-codes}
\end{table}

\vspace{0.1cm}
\noindent
\textbf{Finding \#4: Longevity of API Endpoints.} As already described in finding~\#3, we saw two endpoints going offline during our 2015 experiments. Beyond this, four out of fourteen HTTP endpoints and one out of twelve HTTPS endpoints from our original paper were no longer available when we started our 2018 experiments (cf. Table~\ref{table:changes}). At the time of finishing this paper in May 2020, five out of fifteen API endpoints appear to be unavailable (either permanently or temporarily); two have changed their authentication requirements. We find this observation both surprising and troubling, especially since -- contrary to what one might expect -- there is no apparent relationship between size and popularity of an API and its longevity. All this together indicates that application developers should not necessarily rely on the longevity of the API endpoints used. Instead, they should closely monitor announcements on the provider's websites and have contingency plans for using alternative API endpoints.

\begin{table}[!t]
\center
\caption{Observed Success Ratios of Protocol or Region Change in Case of Failures Across all APIs}
\begin{tabular}{ p{2.3cm}  l  r  r  r }
           
\textbf{Dataset} & \textbf{Strategy} & \textbf{Min} & \textbf{Max} & \textbf{Average} \\
\toprule
2015 (all) & {\scriptsize REGION\_CHANGE} & 7\% & 100\% & 86\% \\
 & {\scriptsize HTTP$\rightarrow$HTTPS} & 0\% & 100\% & 88\% \\
 & {\scriptsize HTTPS$\rightarrow$HTTP} & 7\% & 100\% & 91\% \\
\midrule
2015 (w/o & {\scriptsize REGION\_CHANGE} & 89\% & 100\% & 98\% \\
offline endpoints) & {\scriptsize HTTP$\rightarrow$HTTPS} & 78\% & 100\% & 97\% \\
 & {\scriptsize HTTPS$\rightarrow$HTTP} & 94\% & 100\% & 99\% \\
\midrule
2018 & {\scriptsize REGION\_CHANGE} & 93\% & 100\% & 99\% \\
 & {\scriptsize HTTP$\rightarrow$HTTPS} & 93\% & 100\% & 99\% \\
 & {\scriptsize HTTPS$\rightarrow$HTTP} & 94\% & 100\% & 99\% \\
\bottomrule
\end{tabular}
\label{table:change}
\end{table}

\vspace{0.1cm}
\noindent
\textbf{Finding \#5: Correlation of Unavailability Across Regions and Protocols.} Based on our detailed result logs, we also explored to which degree requests to the same API endpoint experience the same availability behavior when sent (i) from different geographical regions or (ii) via different protocols (HTTP vs. HTTPS). For this analysis, we synchronized the detailed results logs as much as possible -- requests did not start at the exact same time. This means that the actual numbers in the following results should be taken with a grain of salt as they are likely to underestimate the correlation: As requests may in fact be up to five minutes apart, it is certainly possible that there was a short global outage for a period of less than five minutes. Nevertheless, we also saw longer lasting outages for requests from, e.g., a single region, while other regions remained available. So, while the actual numbers are likely to be overoptimistic, the results still indicate that the effect could be leveraged for \emph{some} of the requests. In future work, we plan to explore to which degree this is possible in practice.

\noindent
\emph{How to read the results presented in Table~\ref{table:change}:} The Table reports all aggregated results for different datasets (first column, for 2015 again with and without the two APIs that went completely offline) and strategies. For the strategies, {\scriptsize REGION\_CHANGE} succeeds when one is able to successfully send a request from at least one other region once an API is not available from a specific region. {\scriptsize HTTP$\rightarrow$HTTPS} and {\scriptsize HTTPS$\rightarrow$HTTP} succeed when one is able to successfully send a request via the respective other protocol from the same machine once the first request failed. For each API, we calculated for which percentage of the failed requests the respective strategy would have been successful. The numbers in the table show these success ratios -- for each dataset we show the API with the smallest success ratio (``Min''), with the highest success ratio (``Max''), and the average across all success ratios.

Yahoo! is a good example to explain how regional differences can exist. In their PNUTS paper~\cite{paper_cooper_pnuts}, they describe how Yahoo! rolls out services globally using the underlying PNUTS system for geo-replication across data centers. User requests are then routed to the respectively closest data center and mastership for data records is adapted based on user location. If one data center has availability problems, then all read requests can be served elsewhere. The ability to serve update requests depends on whether the respective master record is stored within the affected data center.

\vspace{0.1cm}
\noindent
\textbf{Finding \#6: General Availability Over the Years.} Overall, availability improved from 2015 to 2018 for all three protocols. Only API-6 got slightly worse (still operating at 99.98\% overall accessibility) and had some problems with ping as described in finding \#2. Ignoring the two offline endpoints from 2015, availability improved from about 99.4\% and 99.38\% (HTTP and HTTPS respectively) to 99.94\% and 99.93\%. See table~\ref{table:overall-availability} which gives an overview of overall accessibility rates across all APIs in 2015 and 2018.

\begin{table}[!t]
\center
\caption{Overall Accessibility for HTTP and HTTPS Requests in 2015 and 2018}
\begin{tabular}{ p{2.3cm} l r r r }
           
\textbf{Dataset} & \textbf{Protocol} & \textbf{Min} & \textbf{Max} & \textbf{Average} \\
\toprule
2015 (all)& HTTP &  32.99\% & 99.992\% & 91.43\% \\
 & HTTPS & 34.17\% & 99.990\% & 93.95\% \\
\midrule
2015 (w/o & HTTP &  94.84\% & 99.992\% & 99.40\% \\
offline endpoints) & HTTPS &  94.79\% & 99.990\% & 99.38\% \\
\midrule
2018 & HTTP &  99.69\% & 99.998\% & 99.94\% \\
 & HTTPS & 99.71\% & 99.993\% & 99.93\% \\
\bottomrule

\end{tabular}
\label{table:overall-availability}
\end{table}

\vspace{0.1cm}
\noindent
\textbf{Conclusion.} Overall, we see the main threat to developers in the longevity of APIs. When we started our experiment in 2015, we did not expect that so many API endpoints would be discontinued three and five years later. We were surprised when endpoints vanished without warning. Also, the correlation of availability across regions appears to be rather low -- this makes it a little bit challenging for an API provider to monitor API availability. Finally, availability appears to have improved from 2015 to 2018 but developers should still be careful in their API selection as not all APIs offer the same availability levels.

\section{Performance Findings\label{sec:latency}}

\begin{figure*}[t]
  \centering
  \includegraphics[width=\columnwidth]{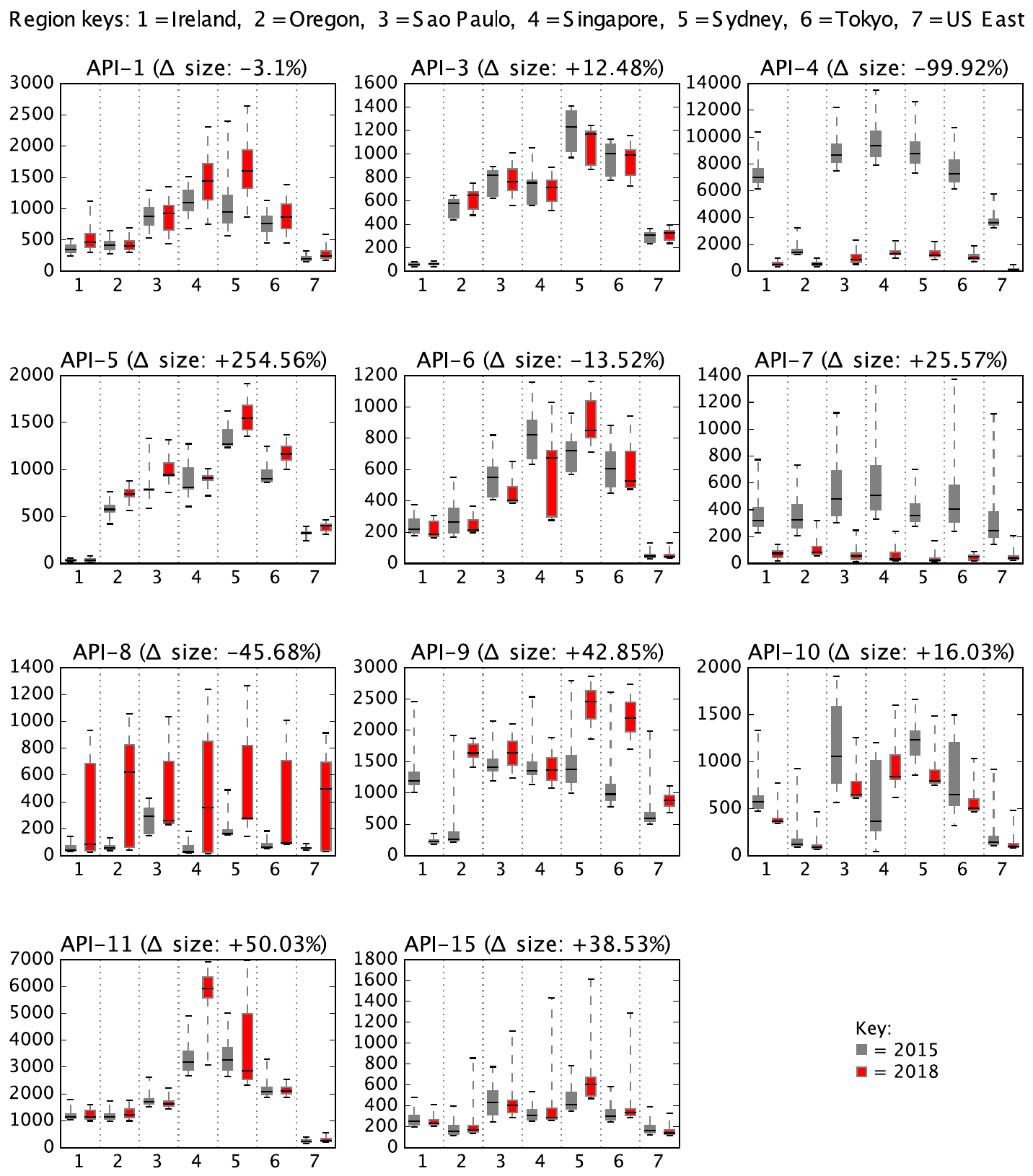}
  \caption{Comparison of 2015 and 2018 HTTPS Request Latency Across Regions in Milliseconds; whiskers mark the 5th and 95th percentiles.}
  \label{fig:latency_comparison_boxplot}
\end{figure*}

In this section, we describe findings of analyzing the latency of web API requests both in 2015 and in 2018, with a focus on the comparison of these findings.
The majority of our findings are reflected in Figure~\ref{fig:latency_comparison_boxplot}, which uses box plots to summarize HTTPS latencies per API and region, comparing results from 2015 (shown in gray) with those of 2018 (shown in red)\footnote{Please, note that the semantics of the whiskers have changed in comparison to~\cite{paper_bermbach_api_benchmarking}. Whiskers now denote the 5th and 95th percentiles which we find more intuitive; originally, we used the default of the Pandas library which is defined as follows: from the top of the box, a space of $1.5*(Q3-Q1)$ (with Q1 and Q3 being the 25th and 75th percentiles) is measured. The largest data point within this space is marked with the end of the whiskers, likewise for the lower whisker.}. We focus on HTTPS to enable comparison with our previous work~\cite{paper_bermbach_api_benchmarking}, and, again, as HTTPS is increasingly becoming the default transport protocol used in the web. Finally, HTTP results do not provide additional insights.

\vspace{0.1cm}
\noindent
\textbf{Finding \#1: Inter-Regional Differences.}
Latency of requests to a single API generally varies significantly depending on the geographic location of the client. We reported this finding in previous work~\cite{paper_bermbach_api_benchmarking} and can confirm this for 2018 as well.
This variance, which we call ``geofactor'', is for a specific API best described by first calculating the mean latency per region $lat^{avg}_i, i\in\{regions\}$. Then the geofactor is calculated as $max(lat^{avg}_i)/min(lat^{avg}_i)$.

In 2015, API-7 had the lowest geofactor at 1.65; in 2018, this honor went to API-8 at 1.47.
In both years, API-5 had the highest geofactor at 28 (2015) and 32 (2018) respectively.
The average geofactor across all APIs increased from 9 in 2015 to 10.9 in 2018.

\begin{figure}[t]
  \centering
  \includegraphics[width=\columnwidth]{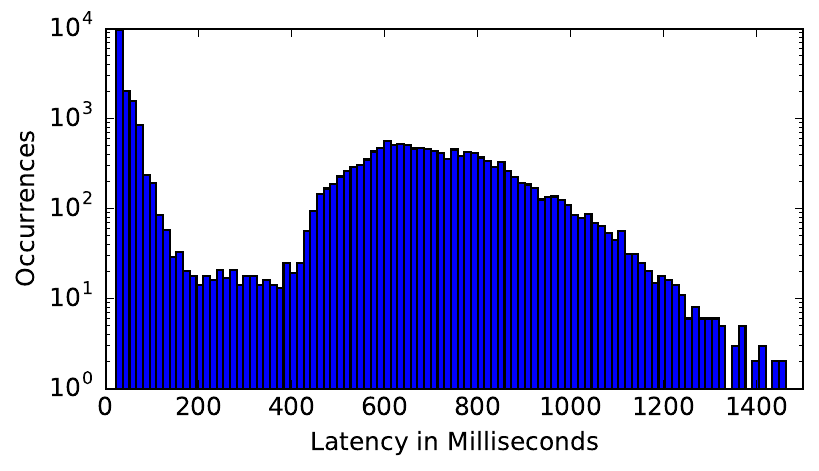}
  \caption{Histogram of HTTPS Latency of API-8 in Ireland in 2018}
  \label{fig:latency_histogram_api8_ireland}
\end{figure}

\vspace{0.1cm}
\noindent
\textbf{Finding \#2: Latency Differences within Regions over Time.}
Even within individual regions, the measured response times of requests to a single API can vary a lot over time.
Consider, for example, Figure~\ref{fig:latency_histogram_api8_ireland}, which shows the histogram of the HTTPS latency measured for API-8 in region Ireland in 2018 (note that the y-axis uses a log scale). As can be seen, a large number of requests finished in 30-150ms. However, another large share of requests requires between 400-1500ms to finish, indicating the vast latency variance in this case.

Furthermore, as reported in our previous work~\cite{paper_bermbach_api_benchmarking}, there are significant outliers in the data reaching response times in the hundreds of seconds, which effectively render these APIs unavailable for those requests.
In the 2018 data, $69$ requests across seven APIs take over one minute to complete -- just over half of these belong to API-1.

\begin{figure}[t]
    \centering
    \includegraphics[width=\linewidth]{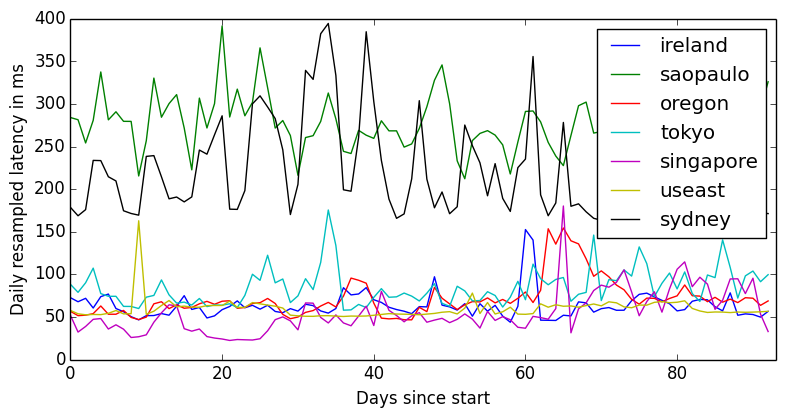}
    \caption{Daily resampled HTTPS latency of API 8 across regions~\cite{paper_bermbach_api_benchmarking}}
    \label{fig:resampled_https_api_8}
  \end{figure}

\begin{figure}[t]
    \centering
    \includegraphics[width=\linewidth]{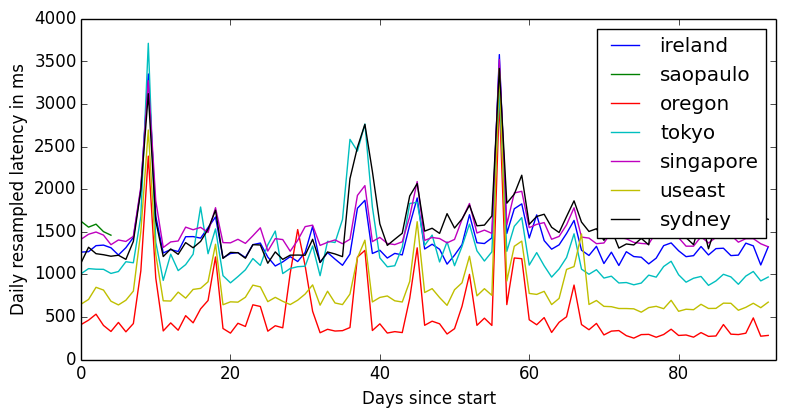}
    \caption{Daily resampled HTTPS latency of API 9 across regions~\cite{paper_bermbach_api_benchmarking}}
    \label{fig:resampled_https_api_9}
  \end{figure}

\vspace{0.1cm}
\noindent
\textbf{Finding \#3: Geo-Distribution of API Endpoints.}
Measuring latency from different geographic locations allows, in some cases, to draw conclusions about the geographic distribution of API servers.
Consider as an example Figures~\ref{fig:resampled_https_api_8} and~\ref{fig:resampled_https_api_9}, which show the average daily latency for API-8 and API-9 respectively across regions.
In case of API-8, latency values fall into two clusters: they are considerably worse in Sao Paulo and Sydney as compared to the rest of the world. Furthermore, the peaks of the regions' latency curves are not aligned, indicating that the API is provided by different servers in different geo locations.
In contrast, in the case of API-9, the peaks of the latency curves of all regions are pretty much aligned, indicating that a single location is responsible for providing that API. Since the measured latency is consistently the lowest in Oregon, it can be assumed that this single API location is closer to Oregon than all other regions.

In the 2018 measurements, API-9 no longer exhibits the described behavior. Rather, latency peaks do not align in a clear way. In fact, in the 2018 measurements, no other API shows latency timeseries that align across all regions.

\vspace{0.1cm}
\noindent
\textbf{Finding \#4: Changes Between Years.}
One way to depict the latency changes per API and region between 2015 and 2018 is to look at the 90\% percentiles. Doing so allows us to delimit the distorting effect that outliers have. We find that, on average across regions, the latency 90\% percentiles increase for 8 APIs, and decrease for 3 APIs.
However, given also the changes in response sizes between the years, individual cases need to be considered: the changes for individual APIs are so different, that aggregated findings across sets of APIs are questionable.
One notable example is API-4, where we had to replace the previously very large resource by a much smaller one, which is reflected in drastically reduced latency in 2018.
Another example is API-7 where, despite the resource size having increased by over $25\%$ between the years, latency has significantly decreased, across all regions. Possible causes for such improvements are the use of caching in the API backend, scaled-up servers to answer requests, or better network speeds.
API-7 is also an interesting example when considering the changes of latency variance between the years. On average across regions, the standard deviation of latencies of API-7 decreased by $88.2\%$.
On the other hand, consider API-8, where the standard deviation grew on average by $193.4\%$ across regions -- an example that shows that increased network speed cannot be seen as the sole factor explaining latency changes between the years.

In summary, we find that multiple APIs denote stark changes in latencies and the variance of latency between 2015 and 2018. This finding underlines once more -- in addition to the latency changes we observed in our individual 3-month experiments -- the necessity for application developers (i) to continuously monitor the qualities of the external services they depend on, and eventually (ii) to deploy mechanisms to mitigate occurring problems.

\vspace{0.1cm}
\noindent
\textbf{Conclusion.}
Measuring latency of web APIs reveals severe implications for application developers. They cannot assume that latency remains constant over time (even on a short time scale, latency can increase by about 10x for a significant number of requests) and have to expect stark latency variance when accessing APIs across the globe (for a given API, the slowest region experiences on average 10x higher latency than the fastest one). If web APIs provide functionalities fundamental to an application, latency variation can have negative impacts on user attention~\cite{Nielsen:1994} or even result in lost business~\cite{Brutlag:2009}. This can be especially expected in cases where latency (temporarily) rises to multiple seconds. Application developers should consider mitigation strategies, as we discussed in previous work~\cite{paper_bermbach_api_benchmarking} and summarize in Section~\ref{sec:discussion_app-engineering}.

\section{Security Findings\label{sec:security}}

In this section, we describe findings from our security experiments. To compare APIs, we have to translate the provider preference list of cipher suites into a numerical score. For this, we define in the following a \emph{cipher suite security score} which describes the security level for a single cipher suite and a \emph{server security score} which aggregates the scores of all cipher suites in the respective provider's preference list.

We define the (heuristic) cipher suite security score $S^{CS}$ per cipher suite as follows:

$$S^{CS} = baseScore + keyLengthModifier$$

\noindent
where $baseScore$ is $1$ if modern algorithms are used, $0$ if no perfect forward secrecy is guaranteed, and $-1$ if a generally known-to-be-weak algorithm (e.g., RC4, DES) is used. The $keyLengthModifier$ is $0.1$ if the selected cipher has a strength of at least 256 bit.

The idea of these scores is not to make absolute statements about the security provided by a suite, but rather to be able to compare servers supporting different sets of cipher suites. For that purpose, based on individual cipher suite security scores, we devise the server security score $S^{S}$ of a server to be:

$$S^{S} = \sum_{i \in CS} \frac{S^{SC}_{i}}{r_{i}}$$

\begin{figure}[t]
  \centering
  \includegraphics[width=\columnwidth]{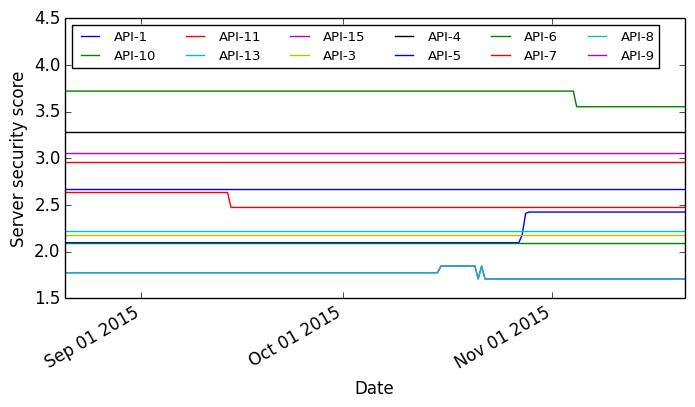}
  \caption{Server Security Score Evolution in Region US East in 2015}
  \label{fig:cipher_score_evolution_useast}
\end{figure}

\noindent
where $CS$ is the ranked list of cipher suites supported by the server (i.e., the ordered list of server preferences), $S^{SC}_{i}$ is the cipher suite security score of a suite $i$, and $r_{i}$ is the rank of the same suite $i$. In other words, the server security score is higher when modern cipher suites are near the top of the ranked list and lower when weak cipher suites are included, especially near the top.

As a toy example, consider a server whose cipher suite preference list is 1.) \texttt{ECDHE-RSA-AES256-SHA384}, 2.) \texttt{ECDHE-ECDSA-AES128-SHA}, and 3. \texttt{RC4-SHA}), which have cipher suite security scores, respectively, of $1.1$, $1$, and $-1$. The resulting server security score would be $(1.1 / 1) + (1 / 2) + (-1 / 3) = 1.267$. Preference list entries starting from the twelfth entry can only affect the second decimal place and beyond. For a preference list with about 30 entries, we can expect the server security score to be in the interval [-4;4].

\vspace{0.1cm}
\noindent
\textbf{Finding \#1: Server Security Scores Differ Across APIs.}
The different APIs feature very different server security scores. Figure~\ref{fig:cipher_score_evolution_useast} shows scores of APIs in 2015 in region US East. As can be seen, API-8 has the lowest score, with an average of $1.76$. On the other hand, API-6 has the highest score, with an average of $3.69$. Similarly, in 2018, average server security scores range from $1.58$ (API-3) to $3.15$ (API-4).

\vspace{0.1cm}
\noindent
\textbf{Finding \#2: Lasting Score Changes in 2015.}
To characterize whether and how security scores of an API change during our experiments, we define a \emph{lasting change} to exist if the security score changes by at least $1\%$ between measurements, and if the new score is maintained for at least ten subsequent measurements (i.e., five days). This definition of a lasting change is informed by exploring plots of the evolution of security scores, which denote some outlying scores and some changes that persist for longer periods (i.e., days/weeks).

\begin{figure}[t]
  \centering
  \includegraphics[width=\columnwidth]{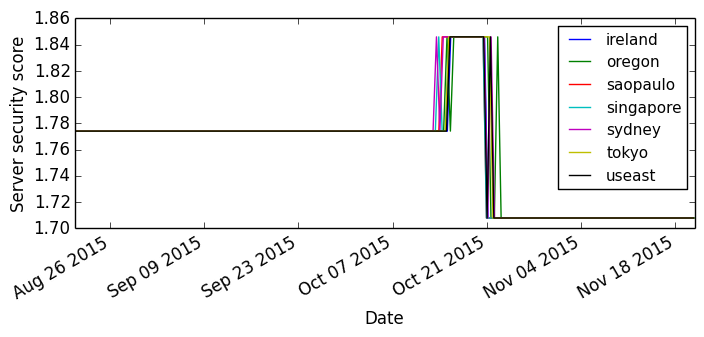}
  \caption{Server Security Score Evolution of API-15 in 2015}
  \label{fig:cipher_score_evolution_2015_api15}
\end{figure}

We find in the 2015 experiments that out of the 12 APIs, 5 (or 41.7\%) denote lasting changes in their security scores (cf. also Figure~\ref{fig:cipher_score_evolution_useast}). Three of these APIs (API-1, API-6, API-7) denote a single change, while the other two APIs (API-8, API-15) denote two lasting changes. In fact, the security scores of of these two APIs nearly perfectly overlap. Information obtained prior to anonymization of our data indicates that these two APIs are deployed on the same managed cloud infrastructure. Figure~\ref{fig:cipher_score_evolution_2015_api15} shows, as an example, the 2015 evolution of the server security score of API-15, which has a (positive) lasting change around Oct 20, and another (negative) lasting change around Oct 26.

Notably, in the 2018 data, we did not find a single lasting change.

\vspace{0.1cm}
\noindent
\textbf{Finding \#3: Changes between 2015 and 2018.}
During our experiments, the server security scores featured lasting changes for only a minority of APIs in 2015, and for no APIs in 2018. However, the server security scores of \emph{all} APIs changed between our two experiments. The median value of the changes (no matter if positive or negative) is $18\%$, with a smallest change of $1.7\%$ (a negative change; API-7) and a maximum change of $27.42\%$ (also a negative change; API-3). In six cases, the score in 2018 is lower than in 2015, in the remaining five cases it is the other way round.

\vspace{0.1cm}
\noindent
\textbf{Finding \#4: Server Security Scores are the Same Across Regions.}
We find that the security scores of APIs were, largely, the same across regions in both 2015 and 2018. I.e., we find no case where security scores were different for long periods of time, or where a lasting change (as defined in Finding \#2) occurred in one region but not in others regions (within at least a few days). The largest time interval between lasting changes appeared for API-15, where the score lastingly decreased from $1.85$ to $1.77$ in region Ireland midday EST of Oct 25, 2015 while the same change occurred in Oregon only midday EST on Oct 27 (i.e., two days later). See again Figure~\ref{fig:cipher_score_evolution_2015_api15}.

\vspace{0.1cm}
\noindent
\textbf{Finding \#5: Discontinuation of Cipher Suites.}
We observe six cipher suites that are only present in the 2015 data. Table~\ref{table:cipher-suites-lost} shows these suites, their cipher suite security score, and the APIs that support them in the 2015 data. Notably, the cipher suite security scores range from -1 to 1.1, i.e., these suites include the whole spectrum of scores (and not only, for example, weak cipher suites).

\begin{table}[t]
  \center
  \caption{Cipher Suites Appearing only in 2015 Experiments}
  \label{table:cipher-suites-lost}
  \begin{tabular}{l c r }
  \textbf{Suite} & $S^{CS}$ & \thead{\textbf{APIs 2015}\\\textbf{with support}} \\ 
  \toprule
  \footnotesize{DHE-RSA-CAMELLIA256-SHA} & 1.1 & 4, 6, 9   \\ 
  \footnotesize{DHE-RSA-CAMELLIA128-SHA} & 1.0 & 4, 6, 9   \\ 
  \footnotesize{CAMELLIA256-SHA}         & 0.1 & 4, 9, 13  \\ 
  \footnotesize{CAMELLIA128-SHA}         & 0.0 & 4 7       \\ 
  \footnotesize{ECDHE-RSA-RC4-SHA}       &  -1 & 1, 8, 10, 15 \\ 
  \footnotesize{RC4-MD5}                 &  -1 & 1, 8, 10, 15 \\ 
  \bottomrule
\end{tabular}
\end{table}

\vspace{0.1cm}
\noindent
\textbf{Finding \#6: Reduced Use of Insecure Cipher Suites.}
Our cipher suite security score assigns a minimum value of -1 to suites that have known vulnerabilities. For example, the RC4 algorithm is known to be vulnerable and was prohibited from use in TLS by the Internet Engineering Task Force in 2015\footnote{For details, cf. \url{https://www.rfc-editor.org/info/rfc7465}}. When counting the frequency in which cipher suites with a score of -1 appear across APIs and across regions, we find 31,316 occurrences in 2015 and only 6775 occurrences in 2018. While this improvement is not reflected in the evolution of server security scores between the years (cf. Finding \#3), it does indicate that over time the ``worst offenders'' are removed.

\vspace{0.1cm}
\noindent
\textbf{Conclusion.}
Quantifying security is a notoriously hard if not impossible challenge. We defined cipher suite/server security scores not to make absolute statements about security of APIs, but to be able to compare API provider security preferences and assess the changes over time. The server security scores of an API are closely related across regions, with only short temporal discrepancy. Also, we only found lasting changes of server security scores in the 2015 experiments.
However, considering the full picture of measurements between 2015 and 2018, we find that all APIs feature changes in security scores, and that these changes are not necessarily leading to higher security. We find, though, that the use of insecure cipher suites has dropped from 2015 to 2018.

\section{Findings from Reaching Out\label{sec:reaching_out}}

In this section, we give an overview of the results of reaching out to all API providers whose APIs we benchmarked in our original paper~\cite{paper_bermbach_api_benchmarking}.

In our assessment, we decided that email or an official contact form is the best way of contacting a provider as it provides (i) a dedicated contact channel (other than a forum that might or might not be monitored) and (ii) a way of communicating potentially sensitive information of the API user. We believe that Twitter and online forums are the second best options as on these channels contact attempts are more likely to be overlooked and also, e.g., in the case of Twitter, require users to entrust potentially sensitive data to a third party. Finally, we ranked all other contact methods worse than this (but better than no contact method). Following this reasoning, we first tried to find an official contact form or support email address for the API. If that was not successful we tried Twitter or online forums. If that also did not work we tried any other contact method that we could find via Google search, LinkedIn contacts of contacts, and other options.

In our contact request, we pointed to our original paper~\cite{paper_bermbach_api_benchmarking} and its findings. We asked the recipient to possibly comment on results, ask questions, or provide us with feedback. We also made it explicit that we intended to cover their response or reaction in this paper unless they would specifically ask us not do this in their response message (we received no such message).

\begin{figure}[t]
  \centering
  \includegraphics[width=0.9\linewidth]{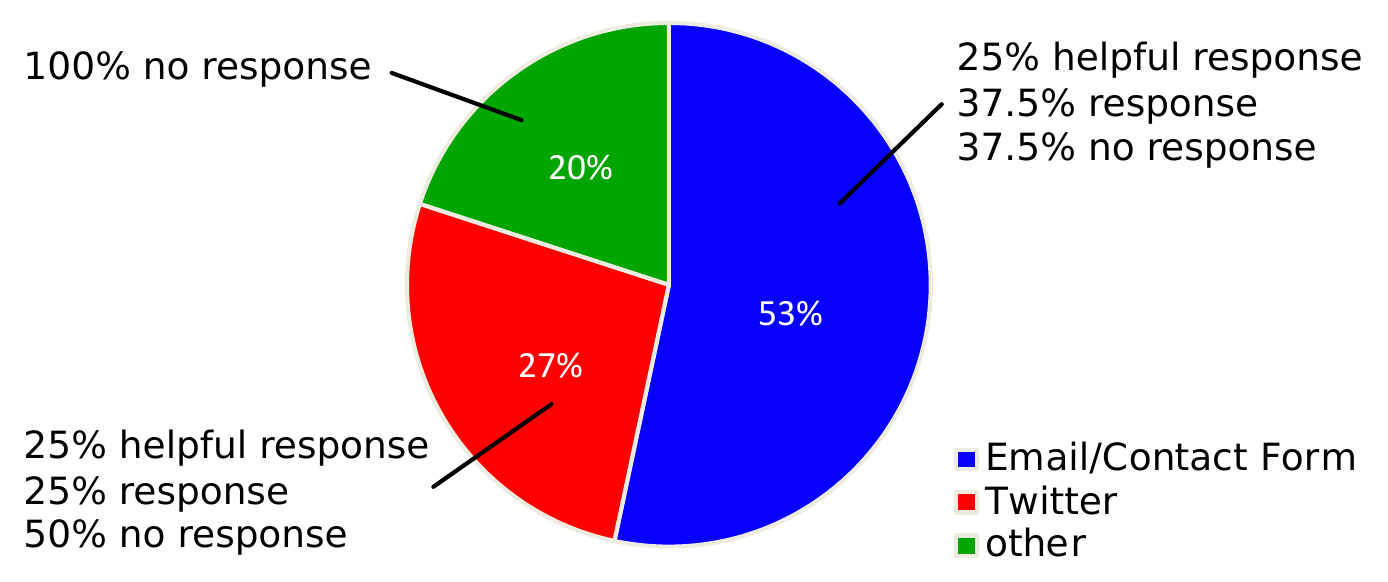}
  \caption{Availability of Contact Methods and Results of Reaching Out to Providers}
  \label{fig:reaching-out}
\end{figure}

\vspace{0.1cm}
\noindent
\textbf{Finding \#1: Availability of Contact Methods.} We were able to find contact forms or official support email addresses for eight (out of fifteen) APIs, four had Twitter profiles. For the other APIs, we contacted one via an existing ``contact of a contact'' on LinkedIn, for one we found a self-proclaimed ``senior developer'' of the API on GitHub and his email address via Google search. Finally, we found one personal website on which the respective person claimed responsibility for that API (and provided some contact details).

\vspace{0.1cm}
\noindent
\textbf{Finding \#2: Quality of Responses (Email/Contact Form).} Of our eight contact requests via official support emails or contact forms, two resulted in interesting conversations where the API provider appeared interested in our results and provided helpful comments. Three of the eight requests resulted in either an automated email or an email that was a bit more personalized which indicated that our request had been forwarded to the respective development teams and that we would get a response soon (we did not). Another three requests were completely ignored.

\vspace{0.1cm}
\noindent
\textbf{Finding \#3: Quality of Responses (Twitter).} Of the four requests we sent via Twitter to the respective API support handles, one provider was very interested and provided helpful comments leading to an email-based discussion. Another provider asked us to send them a private message with further details but we did not receive an answer to that. The last two providers ignored our requests completely.

\vspace{0.1cm}
\noindent
\textbf{Finding \#4: Quality of Responses for Other Contact Methods.} Neither of the three contact requests did result in any reaction of the API provider. See also Figure~\ref{fig:reaching-out} which gives an overview of all responses.

\vspace{0.1cm}
\noindent
\textbf{Finding \#5: Quality of Responses (Summary).} All in all, only three out of fifteen providers offered helpful comments and were actually approachable. There was no correlation with the provider size, country, or industry.

\vspace{0.1cm}
\noindent
\textbf{Finding \#6: Insights Gained.} Aside from the meta-information on whether we received answers or not, we also learned some architectural details. For instance, one of the APIs provides their content strictly read-only. In their deployments, they bundle API code and a database instance. This makes it relatively easy to guarantee high availability and performance as the number of replicas is only restricted by the available budget. Other APIs have it much harder as their backends are subject to the well-known tradeoffs of PACELC~\cite{paper_abadi_pacelc} and other scalability restrictions.

\vspace{0.1cm}
\noindent
\textbf{Conclusion.} Overall, developers should not rely on the ability to reach API providers in the case of problems. Based on our observations, we believe that it is a rare exception when developers can get support unless they are explicitly paying for it. We would recommend to primarily use APIs that are at least well documented and preferably widely used.

\section{Discussion\label{sec:discussion}}

In this work, we set out to repeat experiments originally performed in 2015 and published in previous work~\cite{paper_bermbach_api_benchmarking}.
As such, the scope of our paper is naturally limited by the choices we made in 2015 and the developments on the provider side since then.
In this section, we will discuss possible threats to validity and limitations of our approach as well as overall recommendations.

\textbf{Selection of APIs:}
In 2015, we picked 15 public APIs from a variety of domains, countries, provider types, and popularity levels.
While a larger number may have been preferably, this was not feasible due to the vast amount of data which we needed to analyze using rather time-consuming (manual) exploratory data analysis~\cite{book_bermbach_cloud_service_benchmarking}.
Nevertheless, we believe that our provider selection managed to achieve a large degree of diversity in the set of APIs benchmarked and, hence, covers a broad range of API characteristics that application developers are likely to encounter ``in the wild''.
As such, our experiments also showcase a number of behaviors which application developers have to deal with in practice -- by no means do we wish to insinuate that the list of behaviors observed is representative for the set of \emph{all} APIs or that the list of behaviors is complete.
In fact, the findings from our experiments should not necessarily be generalized to the overall API from which the endpoint stems.
For example, providers may choose to remove an endpoint we benchmarked, while other endpoints of the same API remain available.

Another aspect is that we only used GET requests -- due to caching, actual availability may be worse.
Nonetheless, we deem our results valid examples of how web API qualities can impact applications, which ultimately rely on specific endpoints.
For many APIs, GET may even be the standard way of accessing it, e.g., for a maps API.
We, furthermore, limited our experiments to endpoints that do not require authentication.
One might argue that these endpoints may be of less importance to their providers and may, thus, undergo less scrutiny than other endpoints.
Nonetheless, these endpoints may well be used by applications and the here presented findings should, hence, be considered relevant to application developers.
Still, our observations should not be generalized to endpoints with mandatory authentication in place.

Since our goal was to give a broad overview of behaviors and developments that are observable in practice, reporting anonymized results does not affect reproducibility of results.
In fact, we will gladly provide the pseudonymization mapping on request, e.g., to reproduce our experiments or to correlate our results with external events.
While we are aware that anonymized results render our findings less ``sensational'', we deemed it more important to avoid finger-pointing.

Overall, we believe that we achieved our goal of showing a range of behaviors that application developers can expect in practice as well as their evolution over time.

\textbf{Benchmarking vs. Monitoring:}
Benchmarking is usually defined as creating stress on a system under test while observing its reaction~\cite{book_bermbach_cloud_service_benchmarking}.
Our experiments certainly did not create stress on the API endpoints as the terms of service tend to forbid anything that resembles a DDOS attack.
Monitoring, however, is usually defined as passive observation which we definitely did not do in our experiments~\cite{book_bermbach_cloud_service_benchmarking}.
Overall, we believe that our measurement approach falls into the benchmarking category but is a rather unusual example of benchmarking.

\textbf{Application-Centric Benchmarking:} 
Benchmarking itself is a black box experiment in which a measurement process interacts with a system under test.
Depending on the scenario, additional information may be used in a second step to turn the experiment into a white box or grey box experiment.
The latter category is particularly useful when taking the perspective of providers who aim to improve their offering.

Our goal, however, was to take an application perspective to understand the effects that application developers are confronted with in practice.
As such, our analysis is naturally limited to identifying correlations and deriving possible explanations without the means to verify them through provider-internal knowledge.
In contrast, this also allows us to find results such as the vanishing endpoints which would either not be observed or even disregarded in provider-centric benchmarking.

\textbf{Correction of Previous Results:}
Due to a bug in our benchmarking client, presumably in the JVM runtime used, some measurement threads died and vanished without apparent reason in both 2015 and 2018.
The unexpected death of measurement threads, inexplicably happening without any log entries, meant that data collected in both experiments has a number of missing data points.
As the bug happened without any hint in our logs, which should have captured such an incident, we only discovered this as part of our reanalysis in 2019.
Specifically, manual analysis of the raw collected data reveals that $26$ (or $10.0\%$) of the $259$ measurement threads ($7$ regions, $14$ HTTP, $12$ HTTPS, $11$ ICMP endpoints) died during the 2015 experiments -- dying threads appeared only for HTTP and HTTPS measurements. The death of threads produced no error logs, and subsequent data points were simply missing.
Our original data analysis tool wrongly interpolated these missing data points to mean that the server was unreachable, leading to us over-reporting unavailabilities of APIs in Table~2 of our paper~\cite{paper_bermbach_api_benchmarking}. In Section~\ref{sec:availability} of this paper, we present corrected availability figures, where we no longer attribute missing data points to mean that the API was unavailable.

To avoid such a mishap, we re-wrote our analysis scripts for availability from scratch in Java, thus having an additional implementation (besides our updated Python script that used the pandas library). For both versions of the data analysis code, we wrote additional test cases and compared their outputs to improve their reliability.

\textbf{Implications for Application Engineering:}
\label{sec:discussion_app-engineering}
In our original paper, we discussed that API consumers are directly affected by provider decisions and have no control over the consumed service which was also visible in the results~\cite{paper_bermbach_api_benchmarking}. Namely, latency and availability varied depending on the geo-origin of requests and in general showed a high variance. Also, endpoints vanished during our experiment. As a consequence, we recommended to rely on a variety of mechanisms such as caching, request queuing, monitoring, and API notification services\footnote{e.g., \url{https://www.apichangelog.com}}. We also discussed preliminary ideas for a geo-distributed middleware which acts as a proxy and performs protocol changes or tunnels requests through other regions when necessary and feasible.

Overall, none of these recommendations have become invalid. What is most disturbing, though, is the sheer number of API endpoints that does no longer exist in 2018 -- and neither of these endpoints is for an API that might be simply obsolete such as a service that returns information on outdated technology. Based on this, we can only recommend to actively look for API alternatives (e.g., Google Maps \emph{and} Bing Maps) or to build upon APIs that could be self-hosted if necessary (e.g., Open Street Map). In terms of reaching out, developers should not expect much customer support (this might be better for APIs with a paid plan) so that we would recommend to carefully check existing documentation and developer forums before committing to an API. Finally, security-wise, we have seen that many APIs still support obsolete cipher suites. Since the server selects the cipher suite, we would recommend to restrict the set of supported cipher suites on the client to modern ones only so that downgrades are no longer possible. This also asserts that users are less affected by varying cipher suite preferences of providers.

\section{Related Work\label{sec:relwork}}

In this section, we give an overview of related work starting with web API evolution before describing web API characteristics and benchmarking.

\subsection{Web API Evolution}
Previous work has studied how web APIs evolve, characterizing change patterns and resulting challenges for application developers~\cite{Li:2013}, and assessing how developers react to these changes~\cite{Wang:2014}.
Focusing on implications for applications, Espinha et al.~\cite{Espinha:2015} have shown that mobile applications show diverse behavior in light of web API evolution. Nearly a third of the analyzed applications crash in light of the removal of fields from response messages, highlighting severe consequences that affect user experiences. In a similar vein, Aue et al.~\cite{Aue:2018} report from a single payment API provider's point of view about the cause and scope of erroneous integrations -- reflected by millions of error logs. These papers underline the important role web APIs play for many applications.

More recently, GraphQL~\cite{Wittern:2018,Hartig:2018,Wittern:2019} has evolved as a new paradigm for querying web APIs. Instead of providers defining a fixed sets of endpoints, consumers form queries within the bounds of an API's schema to retrieve or mutate precise subsets of data. This gives API providers more leeway in updating their APIs without breaking client applications as the queries can remain the same. Consequently, we now see wide adoption of GraphQL in industry, e.g., at Netflix~\cite{Shtatnov:2018}, PayPal~\cite{Stuart:2018}, or GitHub~\cite{Torikian:2016}.

In contrast to these papers, we do not address web API evolution, i.e., how the functionality of a web API changes, but the rather focus on web API qualities, i.e., its changing non-functionalities. We argue that these two perspectives complement one another, and can have similar importance. Errors due to faulty integration or issues caused by unavailabilities, security issues, or high latency can equally threaten how users experience applications. Similar, both API providers and consuming developers are challenged to avoid such issues, relying for example on the mitigation strategies hinted at in Section~\ref{sec:discussion_app-engineering}.

Finally, as we could see in our experiments, many availability issues are due to API evolution. While our approach will detect such changes as soon as another benchmark run is started, Yang et al.~\cite{Yang:2018} can detect web API behavior changes based on documentation updates -- if, for instance, such a documentation is maintained in a versioned repository, it is easily possible to subscribe to content changes. Also, Bae et al.~\cite{Bae:2014} can detect API misuse from the application source code and can thus help to avoid issues that result from applications that leverage an unintended loophole in an API which is later on fixed. Both approaches focus on functional behavior.

\subsection{Web API Characteristics and Ecosystems}
A wide array of related work assesses what web APIs exist, how they are used, and how they can be characterized.
Early studies of API ecosystems focused on \emph{ProgrammableWeb}, a community-maintained catalog of web APIs. Analyses have explored the evolution of this ecosystem~\cite{Yu:2008,Weiss:2010}, how APIs in ProgrammableWeb are (reportedly) used in mashups~\cite{Huang:2012}, or utilized such usage relations to recommend web APIs to developers~\cite{Li:2014,wittern2014graph}.
Other studies assessed web API usage outside of ProgrammableWeb, and specifically in the context of mobile applications. For example, Oumaziz et al. perform static analysis on mobile applications to assess if and to what extent they interact with web APIs~\cite{Oumaziz:2017}. Rapoport et al. combine static analysis with a dynamic execution of selected mobile applications, exploring how to best detect consumption of web APIs~\cite{Rapoport:2017}. Finally, Wittern et al. studied how to detect the use of web APIs in JavaScript-based application code mined from GitHub using static analysis~\cite{Wittern:2017b} and a recent study surveyed GraphQL schemas mined from GitHub~\cite{Wittern:2019}.

These publications emphasize the importance of web APIs for application developers and an increasing recognition of related challenges and research opportunities~\cite{Wittern:2017}.

Yet another set of studies assesses characteristics of existing web APIs. 
Rodriguez et al. extract web API requests from large volumes of HTTP traffic logs and assess their adherence to various design principles, including proper use of HTTP verbs or how URLs are structured~\cite{paper_rodriguez_rest_maturity_in_practice}.
Neumann et al. manually identify web APIs among popular web sites and assess their documentations for a similar set of design principles~\cite{Neumann:2018}.
Zdun et al~\cite{zdun2018guiding} study design choices in API design empirically, Ivanchikj et al.~\cite{ivanchikj2018restalk} describe how to derive interaction patterns from API logs, and~\cite{DBLP:conf/europlop/LubkeZPZS19,10.1145/3282308.3282319} model proven API design choices in the form of reusable patterns.

We again see our work to complement these approaches, which focus on design aspects of APIs, as this work focuses on runtime qualities.

\subsection{Benchmarking}
Benchmarking comprises a number of different areas. The oldest area is concerned with quantifying performance of (virtual) machines. These include many SPEC\footnote{\url{https://www.spec.org/}} benchmarks or collections such as the Phoronix Test Suite\footnote{\url{https://www.phoronix-test-suite.com/}} but also more recent developments, e.g., Borhani et al.~\cite{paper_borhani_wpress} who use WordPress as a realistic benchmark application, DocLite~\cite{varghese16} which uses Dockerized light workloads to rank cloud-based virtual machines, or Cloud WorkBench~\cite{scheuner:19-icpe-tutorial,scheuner:14-cloudcom} which automates cloud VM benchmarking. Another well-studied area is the benchmarking of database systems which, in the last few years, has evolved from relational database benchmarks such as the well-established TPC\footnote{\url{http://www.tpc.org}} benchmarks towards modern NoSQL datastores. For instance, the de facto standard YCSB~\cite{paper_cooper_ycsb} and its extensions~\cite{paper_patil_ycsbplusplus,paper_dey_ycsb+t} introduced database benchmarking based on CRUD interfaces, which are more compatible with modern NoSQL stores such as Apache Cassandra~\cite{paper_lakshman_cassandra}. Other approaches, e.g., OLTPBench~\cite{paper_difallah_oltpbench} or BenchFoundry~\cite{paper_bermbach_benchfoundry,paper_bermbach_benchmarking_middleware}, aim to build comprehensive multi-quality benchmarking platforms that also include measurement approaches for qualities beyond performance, e.g., data consistency~\cite{diss_bermbach,paper_wada_consistency_monitoring,paper_anderson_consistency_measurements,paper_zellag_consistency_benchmarking} or elastic scalability~\cite{paper_kossmann_cloud_datastore_benchmarking,paper_kuhlenkamp_vldb,rabl2012solving}. Beyond these, there are a number of approaches studying performance impacts of TLS on NoSQL datastores~\cite{paper_mueller_tls_benchmarking,paper_pallas_et_al_sac_2017_evidence_based,paper_pallas_security_performance_hbase}, web services~\cite{paper_juric_benchmarking_security_webservices}, and web servers~\cite{paper_coarfa_performance_tls_webservers}. Also, Ferme et al.~\cite{ferme2016framework} have studied the performance of workflow systems.

More recently, benchmarking has targeted Function-as-Service, e.g.,~\cite{paper_kuhlenkamp_benchmarking_faas_ccp,paper_kuhlenkamp_faas_foundation,lloyd2018serverless}, has been used in continuous integration and deployment pipelines~\cite{schulz,ferme,van2012kieker,grambow_continuous_benchmarking:_2019,daly2020use}, has been used to evaluate microservices~\cite{paper_grambow_benchmarking_microservices} and their effects on today's data centers~\cite{gan2019open}, and has been used to study capabilities of edge nodes~\cite{mcchesney2019defog}.

Neither of these approaches is directly comparable to the work presented in this paper as these approaches all have in common that they \emph{stress} the system under test while our approach is more \emph{lightweight} and does not create significant load on web APIs as such a practice is explicitly forbidden in the terms of the service of most web APIs. In this regard, our approach is also comparable to monitoring approaches~\cite{paper_kuhlenkamp_aisle} but, in contrast to them, is not entirely passive. For an extended discussion of the broad area of benchmarking -- this list is by no means exhaustive, we refer to our recent book on cloud service benchmarking~\cite{book_bermbach_cloud_service_benchmarking} or foundational publications on the principles of benchmarking such as~\cite{paper_binnig_weather_tomorrow,chapter_gray_database_transaction_processing_handbook,paper_kistowski_building_benchmarks,paper_huppler_building_good_benchmarks,paper_folkerts_cloud_benchmarking}.

To our knowledge, this paper and our original paper~\cite{paper_bermbach_api_benchmarking} are the only publications that aim to quantify quality of service of web APIs through geo-distributed long-term experiments. Other measurement approaches such as artillery.io\footnote{\url{https://artillery.io/}} can be used to run load tests against web APIs. This, however, is explicitly forbidden in the terms of service of all APIs that we have looked at (which includes a much longer list than those used for our experiments). As such, it is unlikely to provide insights into the behavior of web APIs ``in the wild'' or needs to be run by the provider. Finally, one might argue that approaches such as UptimeRobot\footnote{https://uptimerobot.com/} could replace our approach. This is not true since UptimeRobot and other monitoring tools (i) collect and expose only a fraction of the data that we collected and needed, (ii) invoke APIs only from a single location unless the request from that origin failed, and (iii) are neither open source nor extensible. Nevertheless, they are highly valuable tools in practice.

\section{Conclusion\label{sec:conclusion}}

Over the last few years, web APIs have found widespread adoption in mobile, web, and desktop applications. Hence, quality behavior of web APIs more and more affects user experience of such applications -- from a developer perspective, this is highly problematic as they have absolutely no control over third-party APIs and their quality.

In 2016, we published a conference paper on benchmarking of web APIs~\cite{paper_bermbach_api_benchmarking}. For this paper, we extended our original benchmarking tool and repeated the three-month experiment. Of the original 15 APIs from 2015, only 11 remained in 2018. For all these, we reported detailed results and corresponding insights on availability, performance, and provider security preferences. We also analyzed to which degree the API providers can be contacted and actually respond to inquiries.


\clearpage
\bibliography{bibliography}

\end{document}